\documentclass[a4paper]{article}
\pdfoutput=1
\usepackage{icrc2013}
\usepackage[english]{babel}

\title{Cosmic Ray Modulation studied with HelMod Monte Carlo tool and 
comparison with Ulysses Fast Scan Data during consecutive Solar Minima}
\shorttitle{Cosmic Ray Modulation with HelMod}

\authors{P. Bobik$^{1}$, G. Boella$^{2,5}$, M. J. Boschini$^{2,4}$, S. Della Torre$^{2}$, M. Gervasi$^{2,5}$,
D. Grandi$^{2}$, G. La Vacca$^{2,5}$, K. Kudela$^{1}$, S. Pensotti$^{2,5}$, P. G. Rancoita$^{2}$, D. Rozza$^{2,3}$
 and M. Tacconi$^{2,5}$.}
\afiliations
{
$^1$ \textit{Institute of Experimental Physics, Kosice, Slovak Republic}\\
$^2$ \textit{INFN Sezione di Milano Bicocca, I-20126 Milano, Italy}\\
$^3$ \textit{Universit\`a dell'Insubria, I-22100 Como, Italy}\\
$^4$ \textit{Cineca, Italy}\\
$^5$ \textit{Universit\`a di Milano Bicocca, I-20126 Milano, Italy}
\scriptsize{}
}
\email{stefano.dellatorre@mib.infn.it}

\abstract{
The Cosmic Rays propagation was studied in details using the HelMod - 2-D
Monte Carlo code, that includes a general description  of the diffusion
tensor, and  polar magnetic-field. The Numerical Approach used in this work is based on 
a set of Stochastic Differential Equations fully equivalent to the well know 
Parker Equation for the transport of Cosmic Rays.
In our approach the  Diffusion tensor in the frame
of the magnetic field turbolence  does not depends explicitly by Solar Latitude 
but varies with time using a diffusion parameter obtained by Neutron Monitors.
The parameters of the Model were tuned using data during the solar Cycle 23 
and Ulysses latitudinal Fast Scan in 1995. The actual parametrization is able to 
well reproduce the observed latitudinal gradient of protons and the southward shift
of the minimum of latitudinal intensity. The description of the model is also available online
at website www.helmod.org.
The model was then applied on Pamela/Ulysses proton intensity from 2006 up to
2009. The model during this 4-year continous period agree well with both PAMELA
(at 1 AU) and Ulysses data (at various solar distance and solar latitude).
The agreement improves when considering the ratio between this data.
Studies done also with particles with different charge (e.g. electrons) 
 allow us to explain the  presence (or not) of protons and electrons latitudinal gradients  
observed by Ulysses during the Latitudinal Fast Scan in 1995 and 2007.
}


\begin{document}

\maketitle

\section{Introduction}
Galactic Cosmic rays (GCR) entering into the heliosphere experience a 
diffusive process before reaching the inner regions (e.g. Earth orbit). 
This diffusion is mainly due to small scale irregularities of the interplanetary magnetic field (IMF),
generated from the Sun, that permeates and defines the Heliosphere itself. 
During their propagation, GCR are also convected by the expanding solar wind,
they experience a magnetic drift due to the large scale structure of IMF, and finally 
they undergo an adiabatic energy loss. 
The global effect is the cosmic rays flux reduction for
energy below 10$\sim$20 GeV, the so-called \textit{solar modulation}, 
depending on the solar activity, particle
charge and interplanetary magnetic field (IMF) polarity.
Ulysses spacecraft extend the GCR observations outside the ecliptic
(i.e. the plane where the Earth orbit lies) up to $\pm$80$^\circ$ (see e.g. \cite{Heber1996,Simpson1996}).
In this work we present the results obtained by the HelMod Monte Carlo code
(presented in section~\ref{Model}) along the Ulysses orbit during the period
2006 to 2009 and compare the results at Earth location. 
The obtained results were then used to explore the computed latitudinal gradient during
this period and we compared our results with those one obtained during the previous Solar Minimum.

\section{Model Description}\label{Model}
The GCR diffusion process into the heliosphere medium 
with adiabatic energy loss, and the outward convection of solar particles,
were studied in details by Parker in 1965 \cite{parker1965}.
He proposed a transport equation that describes the time evolution of particle 
distribution into the space (e.g. see ref.~\cite{Jokipii77}):
\begin{eqnarray}\label{EQ::FPE_gen}
 \frac{\partial \mathrm{U}}{\partial t}&=& \frac{\partial}{\partial x_i} \left( K^S_{ij}\frac{\partial \mathrm{U} }{\partial x_j}\right) 
- \frac{\partial}{\partial x_i} [ (V_{ \mathrm{sw},i}+v_{d,i})\mathrm{U}]\nonumber \\
&&+\frac{1}{3}\frac{\partial V_{ \mathrm{sw},i} }{\partial x_i} \frac{\partial }{\partial T}\left(\alpha_{\mathrm{rel} }T\mathrm{U} \right) 
\end{eqnarray}
where $\mathrm{U}$ is the number density of Galactic Particles per unit of particle kinetic energy $T$, at the time $t$. $V_{ \mathrm{sw},i}$
is the solar wind velocity along the axis $x_i$~\cite{Marsch2003}, $K^S_{ij}$ is the symmetric part of diffusion tensor~\cite{parker1965}, 
 $v_{d}$ is the drift velocity that takes into account the 
drift of the particles due to the large scale structure of the magnetic field~\cite{JokipiLev1977,Potgieter85,BurgerHatting1995} and finally
$\alpha_{\mathrm{rel}}=\frac{T+2m_r c^2}{T+m_r c^2}$.
The last term of eq.~(\ref{EQ::FPE_gen}) is the adiabatic energy loss ~\cite{parker1965,Jokipii1970}.

This equation can be solved numerically using the HelMod Code (see www.helmod.org \cite{webmodelICRCI2013}).
HelMod Code is based on a Monte Carlo technique to integrate
Parker's equation, in a bi-dimensional (radius and co-latitude)
approximation, from the boundary of an effective heliosphere (in
the present model located at 100 AU) down to the Earth position.

The magnetic drift is included into the model as a convective term 
expressed trough the drift velocity. This is split in
regular drift (radial drift $v_{D_r}$ and latitudinal drift
$v_{D_\theta}$) and neutral sheet drift ($v_{D_{NS}}$), as
described in Ref. \cite{Potgieter85}, and is scaled using the
tilt angle~\cite{wsoWeb} ($\alpha_{\rm t}$) of the neutral sheet
as described in Ref.~\cite{Burger89}. 

In a coordinate system with one axis parallel to the average magnetic field and the other two perpendicular to this
the symmetric part of the diffusion tensor $K^S_{ij}$ is  (see e.g. Refs.~\cite{jokipii1971,DellaTorre2012AdvAstro}):

\begin{equation}\label{eq:Kmag}
 K^S_{ij}=\left [\begin{array}{ccc}
          K_{||} &   0     & 0 \\
           0      & K_{\perp,r} & 0  \\
         0           & 0          & K_{\perp,\vartheta}
        \end{array}\right ] 
\end{equation}
with $ K_{||}$ the diffusion coefficient describing the diffusion parallel 
to the average magnetic field; $K_{\perp,r}$ and
$K_{\perp,\vartheta}$ are the diffusion coefficient describing the 
diffusion perpendicular to the average magnetic field
in the radial and polar directions respectively.
For this work $K_{||}$ is the one proposed by Strass et al.~\cite{Strauss2011} that 
follows from previous study on data taken at the Earth orbit~\cite{Palmer1982,PotgieterFerreira2002,droge2005}:
\begin{equation}\label{EQ::KparActual}
 K_{||}=\frac{\beta}{3} K_0 \frac{P}{1\textrm{GV}} \left(1+\frac{r}{\textrm{1 AU}}\right);
\end{equation}
where  $K_0$ is the diffusion parameter, described in Section 2.1 of~\cite{Bobik2011ApJ}, 
that depends on solar activity and
IMF polarity (as described in Ref.~\cite{DellaTorre2012AdvAstro}),
 $\beta$ is the particle speed in unit of light speed, 
$P=pc/|Z|e$ is the particle rigidity expressed in GV and $r$ is the 
heliocentric distance from the Sun in unit of Astronomical Unit (AU).
As remarked in Ref.~\cite{DellaTorre2012AdvAstro}, in this description $K_{||}$ 
has no latitudinal dependence and a radial dependence $\propto r$, 
nevertheless the frame transformation between the field aligned to spherical heliocentric frame 
(see e.g. Ref.~\cite{burg2008}) introduces a polar angle dependence. 
In Ref.~\cite{DellaTorre2012AdvAstro} we shown how this is sufficient to explain the 
latitudinal gradient observed by Ulysses during the latitudinal \textit{fast scan} 
in 1995 (see e.g. \cite{Heber1996,Simpson1996}).

A complete description of the model and the used parameters can be found in
 Refs. \cite{Bobik2011ApJ,DellaTorre2012AdvAstro}. Further information and results can be found 
in the dedicated website (www.helmod.org). 
Through the website is also possible to obtain the computed proton spectra at Earth orbit every month since 1990.

\section{Modulation outside the ecliptic plane}
Several high precision experiments (e.g. BESS~\cite{SanukiEtAl2000}, 
AMS-01~\cite{AMS01_prot} and PAMELA~\cite{Pamela_Prot}) allow us to 
know in details the modulated proton spectra at Earth orbit for different condition of Solar activity.
Outside the ecliptic plane only the Ulysses spacecraft gave us information on
the heliocentric latitudinal distribution of GCR
 up to $\pm 80^\circ$ of solar latitude at a solar distance from $\sim 1$ up to $\sim 5$ AU.
Proton observation carried out by Ulysses spacecraft during his first \textit{fast scan},
from September 1994 up to August 1995 corresponding to an IMF with $A>0$, show (a) a nearly symmetric latitudinal 
gradient with the minimum near ecliptic plane, (b) a southward shift of the minimum 
and (c) the intensity in the north polar region at $80^\circ$ exceeds the
south polar intensity~\cite{Heber1996,Simpson1996}.
In Refs.~\cite{Bobik2011ApJ,DellaTorre2012AdvAstro} we had shown how the
HelMod code, using the parametrization treated in the previous section,
was able to reproduce both the time variation during the last solar cycle 23
as well the latitudinal distribution of galactic Proton.

The observations performed during the third Ulysses \emph{fast scan}, from 
May to December 2007 corresponding to an IMF with $A<0$,
found an almost zero latitudinal gradient in Proton intensity 
(see, e.g. Ref.~\cite{Heber2008,deSimone2011}). 
In Refs.~\cite{Heber2008} it was estimated that the latitudinal gradient of 2.5 GV
protons ($\sim 1.7$ GeV) is consistent with zero, while Ref.~\cite{deSimone2011} 
founds that for 1.6--1.8 GV Protons ($\sim 1$ GeV) 
the latitudinal gradient during the same period is $(-0.024 \pm 0.005)\%/$degree.
The same study on electron leads to the opposite conclusion:
Ref.~\cite{FerrandoEtAl1996} concludes that, during the $A>0$ \emph{fast scan}, electron intensity 
do not show any evidence of latitudinal dependence, at least up to 2.5 GV.
\begin{figure}[pt]
\begin{center}
\includegraphics[width=0.49\textwidth]{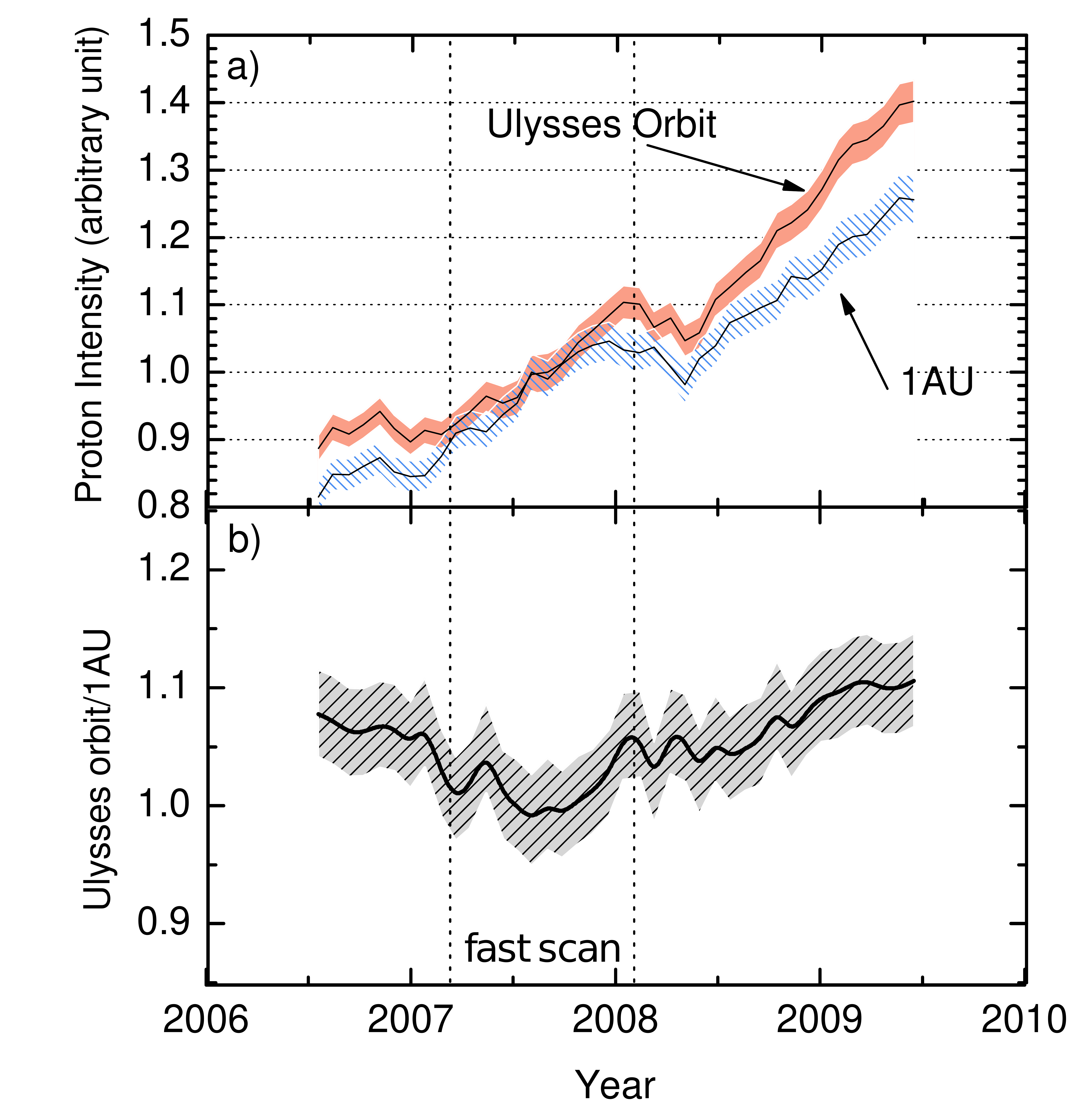}
 \caption{Panel (a): Comparison of HelMod  proton intensities at rigidity$\sim 1.7$GV along Ulysses orbit (red) 
with that at Earth orbit (blue); the proton intensities are normalized to the value corresponding 
to the closest approach of Ulysses spacecraft to Earth orbit. Panel (b): Proton intensity along Ulysses orbit
 divided with those at Earth as computed by HelMod. For both plot the shadow represents the statistical and systematic.
}
 \label{fig:UlyPam}
\end{center}
\end{figure}

In Fig.~\ref{fig:UlyPam}a we show the $\sim 1.7$ GV Proton intensities computed with HelMod Code along 
the Ulysses orbit (red) and at Earth Orbit (blue) in function of time during the 3rd Ulysses \emph{fast scan}.
The data are normalized to the intensity measured when the spacecraft passed 
through the ecliptic plane, i.e. at a closer distance to the Earth.
The HelMod solution was evaluated in the rigidity range 1.67--1.98 GV with a frequency 
of one simulation every 27 day, i.e. a Carrigton Rotation. 
 The blue and red shadow represents the uncertainties of our model, accounting both statistical and systematic
errors as widely explained in Refs.~\cite{Bobik2011ApJ,DellaTorre2012AdvAstro}. 

\begin{figure}[htbp]
\begin{center}
 \includegraphics[width=0.49\textwidth]{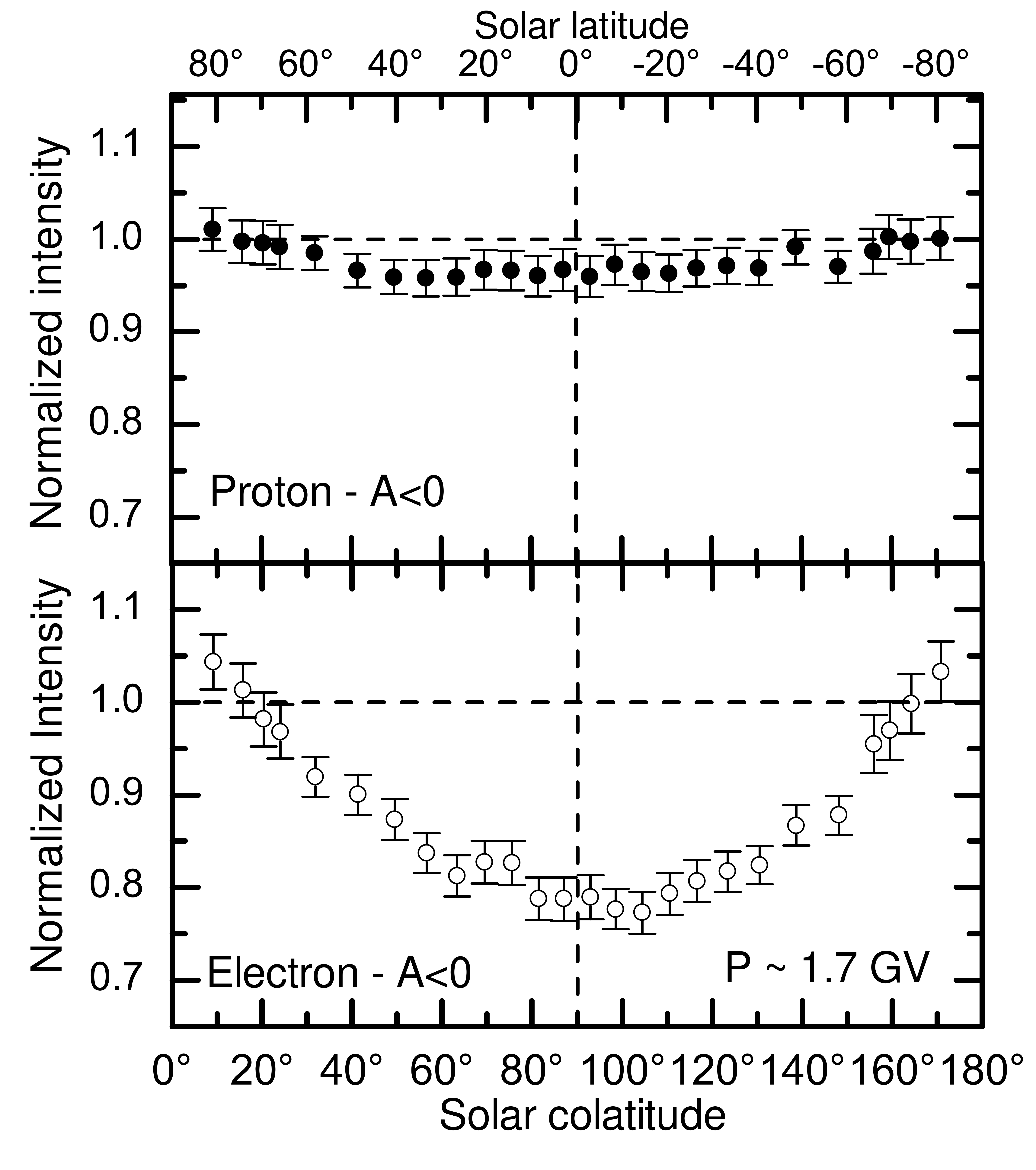}
 \caption{Latitudinal relative intensity along the Ulysses orbit, obtained  at different solar co-latitude
for proton and electron with particle rigidity $~1.7$ GV.
Intensity are divided by the solution at Earth orbit at the same time, then normalized to the average values at 
south pole. Solutions with $A<0$ IMF are evaluated during the Ulysses \textit{fast scan} in 2007.}
 \label{fig:UlyLat2GV}
\end{center}
\end{figure}

Results are compared with those presented 
by Ref.~\cite{deSimone2011} using 
KET/Ulysses and PAMELA (located at the Earth orbit) proton intensity experimental measurements.
It is evident from Fig.~\ref{fig:UlyPam}a how the GCR intensity increases with time so far that
the solar activity decreases.
Simulations outside the ecliptic plane are then normalized for the actual 
solar activity using the intensity at Earth orbit (see Fig.~\ref{fig:UlyPam}b). In this way we get the 
spatial variation of GCR intensity in the heliosphere with radius and latitude.
This technique follow from the one used by Refs.~\cite{Heber1996,Heber2008,deSimone2011}
to study the spatial variation of GCR intensity, and allows us to direct compare with experimental results.
These authors normalized the Ulysses
computed GCR rate with those obtained by near-Earth satellites (like e.g. IMP--8 or PAMELA)
and obtained spatial profile of GCR intensity.
The ratio in Fig.~\ref{fig:UlyLat2GV}b are normalized to be $\sim 1$ at the time of closest approach.
Although in this analysis we do not correct our results for the KET efficiency response as done 
by Ref.~\cite{deSimone2011}, our results are in good agreement with the observed intensities.
This allows us to extend the analysis outside ecliptic plane, as presented in Ref.~\cite{DellaTorre2012AdvAstro}
for $A>0$ period, also to a period with opposite IMF polarity.

In Fig.~\ref{fig:UlyLat2GV} we computed the GCR intensities along the Ulysses orbit.
These are then divided by the intensities computed at the same time at Earth orbit and 
subsequently normalized to average intensity at solar South pole. The GCR intensity
was computed for both protons and electrons during the third  Ulysses 
\textit{fast scan} (i.e. $A<0$ solar minimum period).
From this figure we can conclude, for the studied period, that 
\begin{itemize}
\item the model does not predict a Proton GCR latitudinal gradient, 
\item conversely, a gradient is found reversing the charge of the simulated particle for the same rigidity.
\end{itemize}
The presented results are in a good agreement with conclusion from Ref.~\cite{Heber2008} for the same rigidity interval.
This behavior is a clear consequence of particle charge interaction with the IMF polarity.
The model not including the drift effects shows a small presence of latitudinal gradient of GCR intensity,
but it is not enough to reproduce Ulysses observation in 1995.
The introduction of drift effects create two opposite picture, depending on the product of particle 
charge ($q$) and IMF polarity($A$): with $qA>0$  
the GCR intensity varies strongly with the latitude.
On the contrary, with $qA < 0$ there is a more uniform distribution of GCR intensity.
 This enforces the idea   that the condition $qA > 0$ enhances the entrance of GCR
into the inner heliosphere passing through the poles, while the condition $qA<0$ IMF enhances the GCR arriving from ecliptic region.
This conclusion agrees with similar analysis~\cite{Strauss2012b,FerrandoEtAl1996,Strauss2011}.

\section{Conclusions}
In this work we present the HelMod Monte Carlo Code for the study of Solar modulation.
We extend the application of the Code, that is able to reproduce the complete solar cycle 23, 
outside the the ecliptic region. The qualitative agreements of computed intensities
with those observed by Ulysses spacecraft confirm the statement in literature that the drift mechanism
has a great influence into the GCR propagation in the inner heliosphere. 
The model shows that, in presence of drift mechanism and for rigidities of the order of a few GV, 
the condition $qA>0$ prefers the GCR penetration into the inner heliosphere from high latitude, 
i.e. along the polar regions. On the other side with the condition $qA < 0$ a more uniform GCR distribution with latitude is obtained.

\section*{Acknowledgements}
This work was supported by VEGA grant agency project 2/0076/13.
This work is supported by Agenzia Spaziale Italiana under contract ASI-INFN I/002/13/0, Progetto AMS - Missione scientifica ed analisi
dati.

\end{document}